# Experimental evidence for a two-gap structure of superconducting NbSe$_2$: a specific heat study in external magnetic fields

C. L. Huang,[1] J.-Y. Lin,[2] Y. T. Chang,[1] C. P. Sun,[1] H. Y. Shen,[2] C. C. Chou,[1] H. Berger,[3] T. K. Lee,[4] and H. D. Yang,[1,4*]

[1]Department of Physics, Center of Nanoscience and Nanotechnology, National Sun Yat-Sen University, Kaohsiung 804, Taiwan
[2] Institute of Physics, National Chiao-Tung University, Hsinchu 300, Taiwan
[3] Institute of Physics of Complex Matter, EPFL, Lausanne 1015, Switzerland
[4] Institute of Physics, Academia Sinica, Nankang 11592, Taiwan

To resolve the discrepancies of the superconducting order parameter in quasi-two-dimensional NbSe$_2$, comprehensive specific-heat measurements have been carried out. By analyzing both the zero-field and mixed-state data with magnetic fields perpendicular *(H⊥c)* to and parallel *(H//c)* to the *c* axis of the crystal and using the two-gap model, we conclude that (1) more than one energy scale of the order parameter is required for superconducting NbSe$_2$ due to the thermodynamic consistency; (2) $\Delta_L$=1.26 meV and $\Delta_S$=0.73 meV are obtained; (3) $N_S(0)/ N(0)$=11%~20%; (4) The observation of the kink in $\gamma (H)$ curve suggests that the two-gap scenario is more favorable than the anisotropic *s*-wave model to describe the gap structure of NbSe$_2$; and (5) $\Delta_S$ is more isotropic and has a three-dimensional-like feature and is located either on the Se or the bonding Nb Fermi sheets.

NbSe$_2$ was one of the earliest layered materials known to be superconducting, with $T_c \sim 7$ K, and for 40 years had been thought to be a conventional type II superconductor.[1,2] Moreover, the charge density wave (CDW) occurring at 33 K in NbSe$_2$ is also intriguing,[3] and was once proposed to originate from a combination of the fractional nesting of Fermi surfaces (FSs) around $K$ and the saddle band, and was related to the anomalies of FS attributed to a quasi-two-dimensional (quasi-2D) structure with hexagonal symmetry.[4,5] Because the superconductivity and CDW compete with each other for FS, consequently, the interplay between superconductivity and CDW in NbSe$_2$ has been abundantly investigated[1-6] and the relation between Fermi surface nesting and CDW is still under debate in current literature.[7] On the other hand, due to the recent discovery of new superconductors, new concepts of the superconducting order parameter have been well established, such as $d$-wave pairing in cuprates[8] and two-gap superconductivity in MgB$_2$.[9,10] Since then, interest has revived in the superconducting order parameter of NbSe$_2$.[4,11-22] Recent studies of the superconducting order parameter in NbSe$_2$ generally agree that more than one energy scale is involved.[4,11-22] Beyond that, research is inconsistent in terms of the magnitude of the superconducting gaps.[11,12,16] In particular, there also remains an even more heated controversy on which Fermi sheets the large and small gaps are located, respectively.[19] In this study, we have carried out comprehensive specific-heat measurements both at the zero magnetic field and in the mixed state to resolve these discrepancies. Being a bulk probe, low-temperature specific heat has been recognized as one powerful tool to determine the pairing state in superconductors and gives additional information on the vortex state.[8-10] The previous specific-heat reports on NbSe$_2$, though showing part of the data in the present work, did not fully analyze the issue of multigap or anisotropic superconductivity.[13-15,18] In this work, the magnetic field $H$ is applied both perpendicular $(H \perp c)$ to and parallel (*H*//*c*) to the *c* axis of the crystal, respectively, to study the anisotropy of the order parameter. The present thermodynamic approach leads to a two-gap scenario with $\Delta_L$=1.26 meV and $\Delta_S$=0.73 meV in the two-gap model. The small gap $\Delta_S$ is somewhat isotropic and, therefore, is either on the Se or the bonding Nb Fermi sheets.

The single crystalline dichalcogenide NbSe$_2$ was grown by the standard iodine vapor transport method.[23] Stoichiometric amounts of 99.9% pure Nb and 99.999% pure Se shots were sealed in a quartz ampoule and then heated in a temperature gradient for several weeks. The low-temperature specific heat *C*(*T*,*H*) was measured with a $^3$He heat-pulsed thermal relaxation calorimeter in the temperature range from 0.6 to 10 K by applying different magnetic fields with *H*⊥*c* and *H*//*c* to the *c* axis of the single crystal. Details of the experiments are described in Refs.10 and 24.

Figure 1 shows the result of the specific heat *C*(*T*,*H*) of NbSe$_2$ plotted as *C*/*T* vs. *T*$^2$ with *H* from 0 to 8 T. With decreasing *T*, the specific-heat jump starts at *T*~6.9 K, consistent with that in another specific-heat study.[15] The thermodynamic *T$_c$*=6.70 K is determined by the local entropy balance around the phase transition anomaly. At zero field, *C*(*T*)/*T* approaches a zero interception at low *T*, indicating a complete superconducting volume and the high sample quality. For *H*//*c*, there is no observed superconducting transition anomaly at *H*=5 T. Thus the normal-state specific heat *C*$_n$(*T*) is extracted from *H* = 5 T data between *T* = 0.6 and 10K with *H*//*c*. *C*$_n$(*T*) can be described as *C*$_n$(*T*) = *γ*$_n$*T* + *C$_{lattice}$*(*T*) where *γ*$_n$*T* is the normal electronic contribution and *C$_{lattice}$*(*T*) = *β*$T^3$+*α*$T^5$ represents the phonon contribution. The fit leads to *γ*$_n$=19.09 ± 0.20 mJ/mol K$^2$, *β*=0.517 ± 0.007 mJ/mol K$^4$ (corresponding Debye temperature Θ$_D$=224 K), and *α*=(2.7 ± 0.6)×10$^{-4}$ mJ/mol K$^6$.

The superconducting electronic contribution *δC*(*T*) can be obtained by *δC*(*T*)=*C*(*T*,*H*)-*C*$_n$(*T*), and *δC*(*T*)/*T* vs *T* for *H*=0 is shown in Fig. 2. The entropy conservation required for a second order phase transition is fulfilled, as shown in the inset of Fig. 1(b). This check verifies the thermodynamic consistency of the measured data. By the balance of entropy around the transition, the dimensionless specific-heat jump $\delta C/\gamma_n T_c$=2.12 at $T_c$ is determined, which was also observed and emphasized in a study by Sanchez et al.[13] This value is much larger than the weak coupling limit value 1.43 and, thus, indicates a moderate or strong coupling scenario in NbSe$_2$. To elucidate the superconducting order parameter, $\delta C(T)/T$ is analyzed by the isotropic *s* wave, line node order parameter, anisotropic *s* wave and the two-gap model. The order parameters used to fit the data are $\Delta=\Delta_0$ for isotropic *s* wave, $\Delta=\Delta_0 \cos3\phi$ for line nodes and $\Delta=\Delta_0(1+\alpha \cos6\phi)$ (where $\alpha$ denotes the gap anisotropy) for anisotropic *s* wave. In the two-gap model, two (assumed) isotropic order parameters $\Delta_L$ and $\Delta_S$ are introduced as in the previous works.[9,25,26] It is clear that neither isotropic *s* wave nor line node order parameter is consistent with the experimental data. On the other hand, both anisotropic *s*-wave and the two-gap model could result in satisfactory descriptions of the data. To achieve optimal fitting, the anisotropic *s*-wave model requires that $\alpha$=0.35 and $2\Delta_0/kT_c$=4.2. The two-gap model with $2\Delta_L/kT_c$=4.5, $2\Delta_S/kT_c$=2.6 and $\gamma_{nL}/\gamma_{nS}$=80%:20% also leads to an accurate description of experimental data. In order to further distinguish between the anisotropic and two-gap model for the present case, magnetic field dependent data in the mixed state are analyzed and the Fermi surfaces of NbSe$_2$ are discussed.

In Fig. 3, $\gamma(H)$, obtained from the linear extrapolation of the low-*T* data in Fig. 1 to *T*=0, is shown for *H//c* and *H*⊥*c*, respectively. The fitting temperature range is $T^2$≤2 K$^2$ for data of all the fields in both directions, except for *H//c*=4 T in which the superconductivity suffers nearly complete suppression, and data below *T*=1 K are analyzed. The example of $\gamma(H)$ determination for *H*⊥*c* is shown in Fig. 1(c). Due to the page limit, the case of *H//c* is not shown here. Previously, $\gamma(H)$ was thought to mainly result from the quasiparticle contribution inside the vortex cores and, therefore, $\gamma(H)$ was linear with respect to $H$ for $s$-wave pairing.[8] However, recent studies of the mixed state, which include the contribution of the delocalized quasiparticles and the factor of the reduced vortex core size with increasing $H$, suggest that $\gamma(H)$ is nonlinear with respect to $H$ for even the simplest isotropic $s$-wave case.[27-30] Furthermore, there does not exist a simple power law relation between $\gamma$ and $H$ for the whole range of $H_{c1}<H<H_{c2}$.[31-32] Recent progress on the numerical calculations in the mixed state has allowed a semiquantitative analysis of $\gamma(H)$ for different order parameters.[26-29] In the following, we utilize the numerical results from Nakai *et al.*[30] to analyze the present experimental $\gamma(H)$. Numerical results from different works may deviate slightly; however, the main conclusions in the present work would not be altered by such a deviation. The numerical result for isotropic $s$ wave was plotted as the dashed line in Fig. 3(a).[30] The different order parameters $\Delta$ can be chosen for a comparison with the experimental results as shown in Fig. 3(a). First, the observed $\gamma(H)$ is inconsistent with the simple isotropic $s$-wave model. Second, the anisotropic $s$-wave model fits $\gamma(H)$ reasonably, and is consistent with the results in Ref. 30, yet the obtained anisotropy value of $\alpha=0.5$ is larger than $\alpha=0.35$ from the zero field data. Third, the fit of the two-gap model (standard deviation, SD = 0.112, where SD=$[\Sigma(\gamma_{exp}-\gamma_{model})^2/N]^{1/2}$ and N is the number of data points) actually leads to a better fit, with a smaller standard deviation than that by the anisotropic $s$-wave model (SD=0.322). To more clearly show the difference, we have plotted DF in the inset of Fig. 3(a). The fit of the two-gap model even reproduces the small kink (or say, change in slope) around $H$=0.75 T, which is attributed to the upper critical field $H_{c2S}$=0.62±0.06 T associated with the small gap $\Delta_S$. The fit also leads to $\gamma_{nL}/\gamma_{nS}$=82%:18%, almost identical to the ratio from the zero-field data. In this sense, the two-gap model is better than the anisotropic $s$-wave in describing the in-field data. In Fig. 3(b), $\gamma(H)$ is shown for $H//c$ and $H\perp c$. Note that $\gamma(H)$ for $H\perp c$ also shows a kink around $H$=0.75 T. Indeed, the $H\perp c$ data can be described well by $H_{c2S}$=0.73±0.10 T and $H_{c2L}$=15.2±0.3 T. It is interesting to note that $H_{c2S}$ is almost the same for both $H//c$ and $H\perp c$. This infers that $\Delta_S$ is more isotropic and rather three-dimensional. Moreover, the anisotropy parameter $\Gamma = H_{c2L,H\perp c}/H_{c2L,H//c}$=15.2/4.9=3.1 is identical to the value from the torque measurements,[33] again strongly indicating the consistency of the two-gap model both in zero- and in-field data.

Further support for the two-gap model can be found in that the anisotropic s-wave model would not give a kink observed in the data. In order to explore the possible kink in $\gamma(H)$, the evolution of $C(H)/T$ with $T$ is depicted in the inset of Fig. 3(b). It clearly demonstrates that the kink structure at low $T$ is rapidly smoothed out with increasing $T$. This is the most likely explanation for the lack of a kink in $C(H)/T$ (at $T$=2.3 K) of Ref. 15. Further information related to the kink in $\gamma(H)$ is provided in the inset of Fig. 1(a). The slope of the linear fit changes abruptly from $H$=0.5 T to $H$=1 $T$, in contrast to the smooth increase in the slope from $H$=1 to 8 T. This abrupt change in the same field range was also observed for $H//c$ (not shown). Qualitatively, the slopes from the linear fit indicate the increase in $C(T)$ from the quasiparticle contribution with increasing $T$. The abrupt increase in the slope for 0.5 T<$H$<1 T is, thus, attributed to the suppression of the small gap in some portion of the Fermi surface. It is noted that the above value of $H$ is consistent with that for the kink in $\gamma(H)$. Moreover, it is noted that thermal conductivity experiments also observed a slope change in the plot of $\kappa/T$ vs. $H$ at about $H$=0.6 T, which was explained in the context of multiband superconductivity.[4] Actually, a similar kink to that of the present work can be found in the specific-heat data of Refs. 4 and 14 combined as $H//c$ [see Fig. 3(a) in Ref. 4], as well as with the thermal conductivity data $\kappa(H)$ of Ref. 34 as $H\perp c$. In Ref. 34, a phase transition occurred at $H$ ~ 10 kOe in the hysteretic behavior of $\kappa(H)$, which may be related to the origin of the kink, and which of those are possibly due to gapless FS.[22] Therefore, the kink structure probably reflects the nature of the low $T$ data.

The above data and analysis establish that more than one energy scale is needed to explain the specific-heat data. Moreover, the small energy gap is three-dimensional-like. The issue remains on the assignment of the two simplified energy gaps to various Fermi sheets in $NbSe_2$. This point can be further illustrated with the help of a basic picture of the Fermi surfaces as shown in Fig. 4.[17] It denotes two Nb-derived bands and one Se-derived band crossing the Fermi energy. In total, there are five distinct Fermi sheets. The four Nb-derived Fermi sheets are quasi-2D, and the Se-derived Fermi surface is in the shape of a three-dimensional (3D) pancake. It seems natural to assign $\Delta_S$ to the Se-derived Fermi sheet, and $\Delta_L$ to the other four Nb-derived quasi-2D Fermi sheets. Actually, $\Delta_L$=1.26 meV is consistent with the observed energy gap value of 1.13 meV~1.22 meV on the Nb-derived sheets by angle resolved photoemission spectroscopy (ARPES).[16] However, the low energy scale of the gap $\Delta_S$=0.73 meV (0.7 meV by scanning tunneling spectroscopy[11] and 0.6 meV by the de Hass-van Alphen experiments[12]) was not observed by ARPES (partly due to the energy resolution and the low $T$ limit of ARPES). The derived gap values are fairly consistent with tunneling spectroscopy results,[20,21] where a highly anisotropic (or multiband) gap with $\Delta_{max}$ = 1.4 meV and $\Delta_{min}$ = 0.4~0.7 meV was revealed. Related to $\Delta_S$, the present specific-heat data require that the Fermi surface density of states (DOS) $N_S(0)$ should be 11%~20% of the total DOS $N(0)$. Density functional theory (DFT) calculations suggest $N_{Se}(0)/N(0)$~5% in Ref. 7. Nevertheless, $N_{Se}(0)$ can be effectively enhanced to the observed values by only a minor charge transfer (≈0.04 electrons) from the Nb bands to the Se band in DFT calculations, due to the large effective mass of the latter.[7] However, if the renormalization factor in $\gamma$ is considered, a larger number of the charge transfer is required. On the other hand, it is noted that de Hass-van Alphen measurements suggest a Se pancake which is considerably smaller than the local density approximation estimates.[12] A recent work based on the penetration depth study also reported anisotropic or two-gap superconductivity in NbSe$_2$.[19] That work suggests that the energy gap on the Se-derived sheet is at least as large as that on the Nb-derived sheets, and the small energy gap is on some portion of the Nb-derived sheets. This is not consistent with the relatively uniform energy gap on the Nb derived sheets observed by ARPES.[16] However, the penetration depth study's results might find some agreement in the present study. The bonding Nb (BN) Fermi sheets have strong warping along $k_z$ and are estimated to have ~40% of $N(0)$. Actually, the average $v_{Fx}/v_{Fz}(\approx 2)$ implies that BN sheets are not extremely anisotropic.[7] Following the context of Ref. 19, $\Delta_S$ is likely on some portion of the BN sheets or the superconducting energy gap on BN sheets is anisotropic. To further clarify this issue, the more comprehensive ARPES works on NbSe$_2$ are necessary.

To summarize, the comprehensive magnetic field dependence of specific-heat measurements has been performed to study the superconducting order parameter of NbSe$_2$. The two-gap model with $\Delta_L$=1.26 meV, $\Delta_S$=0.73 meV, and $N_S(0)/N(0)$=11%~20% can accurately describe the gap structure. By analyzing the in-field data with $H//c$ and $H\perp c$, respectively, we argue that the two-gap scenario is more favorable than the anisotropic $s$-wave model in describing the gap structure of NbSe$_2$. Also, the small gap $\Delta_S$ is more isotropic and has a 3D-like feature and is located either on the Se-derived Fermi surface or on the bonding Nb Fermi sheets. The present two-gap scenario largely clarifies the controversies and debates over the order parameter of superconducting NbSe$_2$ in the previous literature.

We are grateful to A. P. Ramirez and N. Nakai for the issues in the mixed state, and to I. I. Mazin and Y. R. Chen for the discussions on the electronic structure of NbSe$_2$. This work was supported by National Science Council of Taiwan under grant Nos. NSC 95-2112-M110-023 and NSC 95-2112-M-009-036-MY3, and MOE-ATU projects.

**Figure Captions**

FIG. 1 (color online). Temperature ($T$) dependence of specific heat ($C$) plotted as $C/T$ vs. $T^2$ at various magnetic fields for NbSe$_2$ in (a) $H \perp c$ (from right to left: $H$ = 0, 0.3, 0.5, 1, 1.5, 2, 3, 4, 6, and 8 T) and (b) $H // c$ (from right to left: $H$ = 0, 0.1, 0.2, 0.3, 0.5, 0.75, 1, 2, 3, 4, and 5 T). Insets: (a) The determination of $\gamma(H)$ from the linear extrapolation of data for $T^2 \leq 2$ K$^2$ and (b) entropy conservation for superconducting phase transition.

FIG. 2 (color online). Various fitting of $\delta C(T)/T$ vs. $T$ using (a) isotropic $s$-wave, (b) line node, (c) anisotropic $s$-wave, and (d) two-gap models. The deviation of each fitting from data is shown in the respective inset, where DF represents the difference between the fitting values and actual data.

FIG. 3 (color online). $H$ dependence of $\gamma(H)$ for (a) $H//c$ and (b) $H \perp c$. The error bar of the data is roughly the size of the diameter of the circle. Inset: $C/T$ vs $H$ at $T$=0.62, 0.72, 1.1, and 2.3 K, respectively (with offsets). Arrows indicate the kinks in low $T$ data. The DF, which represents the difference between $\gamma_{\text{exp}}$ and $\gamma_{\text{model}}$, is shown in the inset, where the solid red circles are for the two-gap model and the hollow circles are for the anisotropic $s$-wave model.

FIG. 4 (color online). The Fermi surfaces of NbSe$_2$ and the associated superconducting order parameters. The shapes of the five distinct Fermi sheets mainly follow Ref. 17. See text for the discussion on the two alternative scenarios.

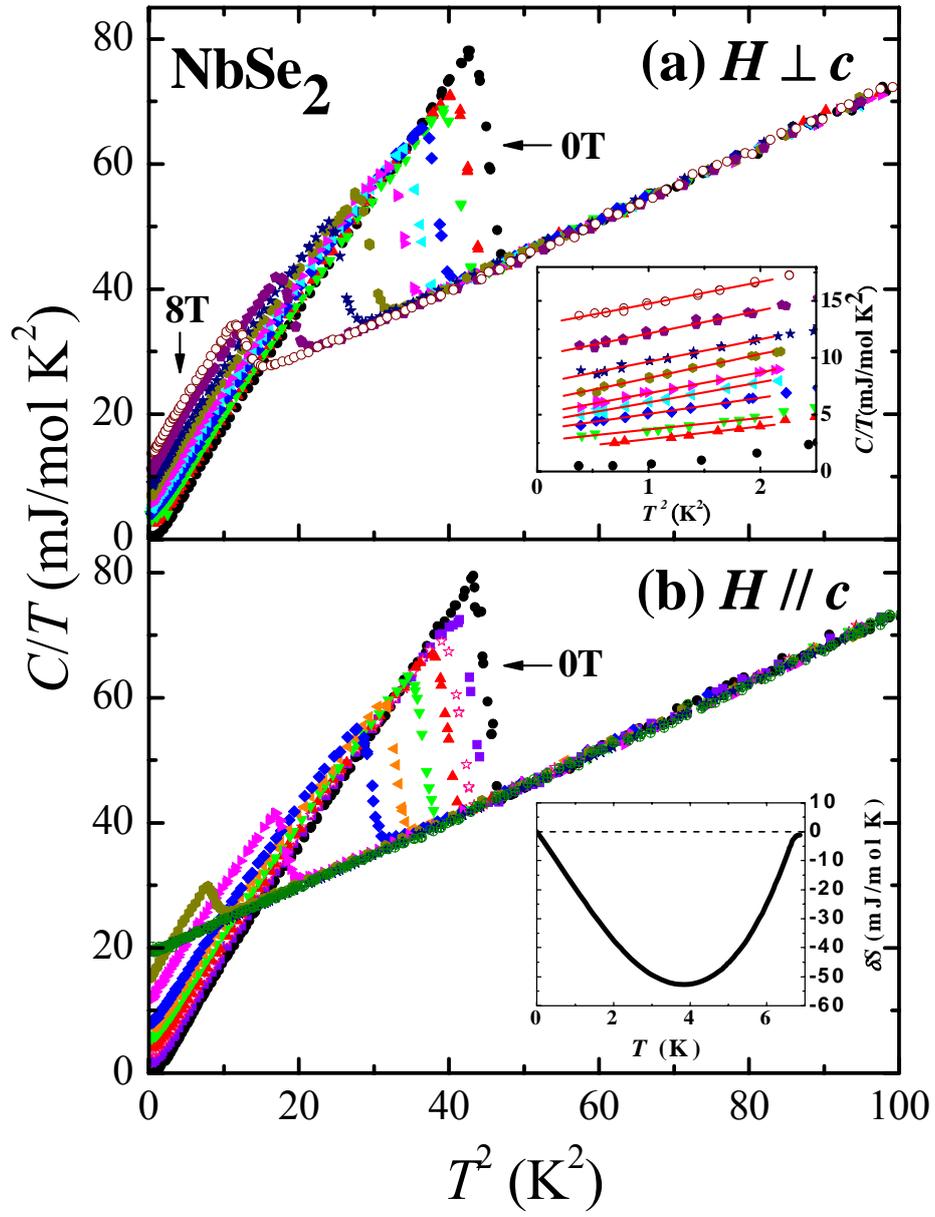

Fig. 1  Huang *et al.*

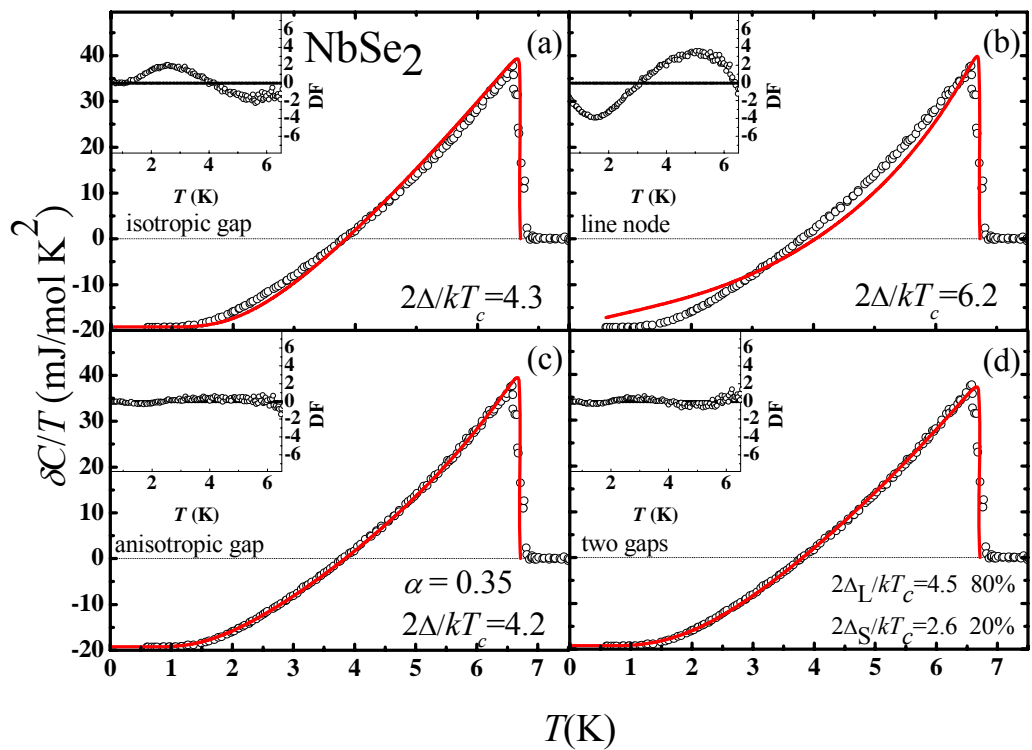

Fig. 2  Huang *et al.*

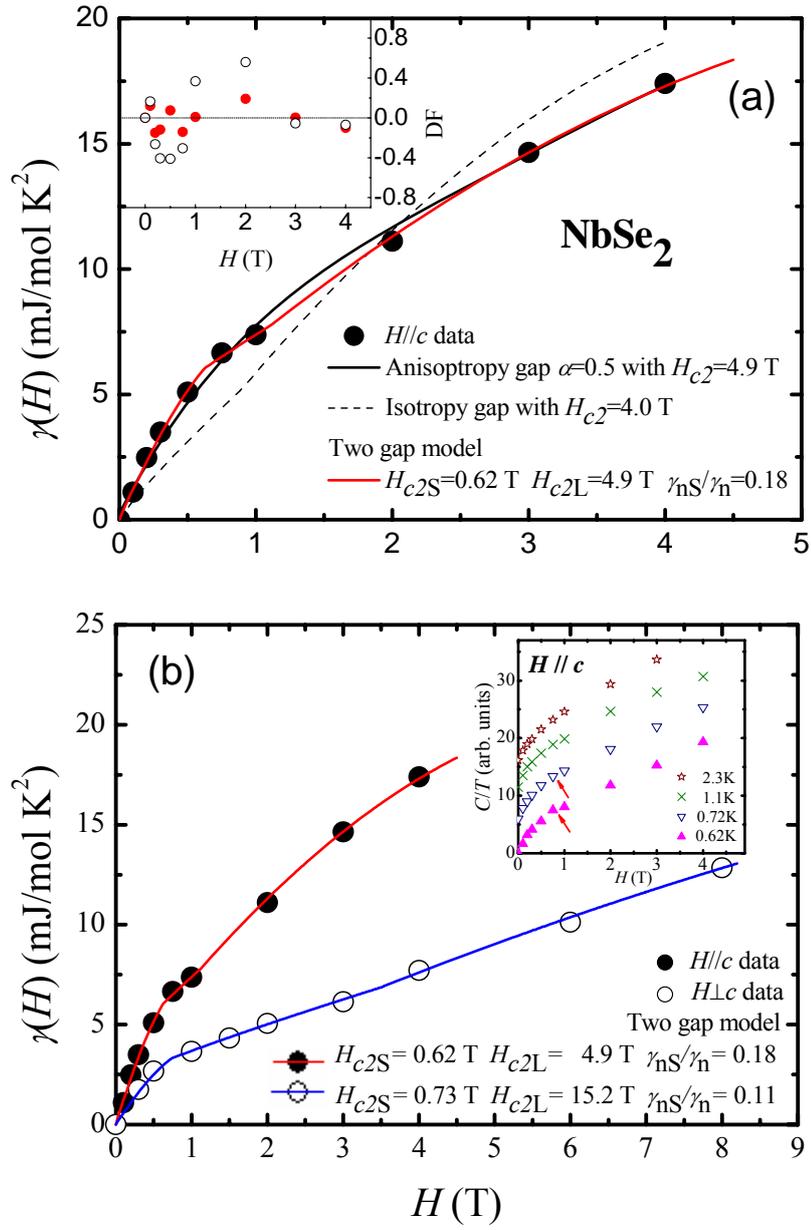

Fig. 3   Huang *et al.*

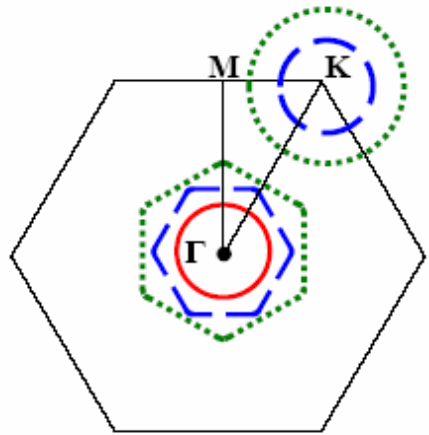

Fig. 4   Huang *et al.*